\documentclass{vienna-conf2019}
\usepackage{graphicx}
\usepackage{hyperref}
\usepackage[]{natbib}
\usepackage{epstopdf}

\def\BibTeX{{\rm B\kern-.05em{\sc i\kern-.025em b}\kern-.08em
    T\kern-.1667em\lower.7ex\hbox{E}\kern-.125emX}}
\bibpunct{(}{)}{;}{a}{}{,}  %%%%%%%%%%%%%  A&A bibliography style
\begin{document}

\TitreGlobal{Stars and their variability observed from space}

%%-----------------------------------------------------------------
%%      the top matter
%%

\title{Binary stars: a cheat sheet}

\runningtitle{Binaries as key laboratories for stellar physics}

\author{J. Southworth}\address{Astrophysics Group, Keele University, Staffordshire, ST5 5BG, UK}

%% IF Author3 has the same affiliation than Author1:
%\author{C.\,E. Author3$^1$}

%% IF Author3 has its own affiliation:
%\author{C.\,E. Author3}\address{Dept. of Chess, University of Games, 35101 Las Vegas, Monaco}

%% IF Author3 has two affiliations, the one of Author1 and a second one:
% \author{C.\,D. Author3$^{1,}$}\address{Dept. of Chess, University of Games, 35101 Las Vegas, Monaco}

%% Keep this line, even if the page will be settled afterwards.
\setcounter{page}{237}

%%-----------------------------------------------------------------

\maketitle

%%-----------------------------------------------------------------
%%        The abstract
%%
%%  Warning!  within the abstract:
%%  - do not use macros.
%%  - do not use commands like: \cite, \citet, \citep ... etc.

\begin{abstract}
I present a brief summary of three different types of binary star -- astrometric, spectroscopic and eclipsing -- and tabulate the properties of these systems that can be determined directly from observations. Eclipsing binary stars are the most valuable of these, as they are our main source of direct mass and radius measurements for normal stars. In good cases, masses and radii can be obtained to better than 1\% precision and accuracy using only photometry, spectroscopy and geometry. These measurements constitute vital empirical data aginst which theoretical models of stars can be verified and improved. I give examples of the use of these systems for constraining stellar theory and the distance scale, and conclude with a presentation of preliminary results for the solar-type eclipsing binary 1SWASP J034114.25+201253.5.
\end{abstract}

%% Insert the keywords (to appear in the ADS indexing)
%% Keywords must be separated by a comma
%% They must be chosen from the keyword list of A&A

%%-----------------------------------------------------------------

\section{Introduction}

Binary stars are one of the classical subject areas of \emph{astronomy}, as they represent the only way of determining the masses and radii of normal stars to high precision and accuracy. This makes them \emph{astrophysically} vital: the properties of stars in binary systems are used to calibrate theoretical stellar models, determine the distances to nearby galaxies, and support asteroseismological studies.

In this review I summarise the various types of binary star, the history of their study, and what physical properties can be obtained from them. I then present some new work in progress on the eclipsing binary 1SWASP J034114.25+201253.5, detected using SuperWASP data %\citep{Pollacco+06pasp}
and observed using the NASA K2 mission. %\citep{Howell+14pasp}.

\section{Binary stars}

Table\,\ref{tab:par} summarises the properties directly measurable for different types of binary star. Both their symbols and names are given. The organisation of the table follows the observational techniques used: a convenient way of considering these systems.

\subsection{Visual binaries}

The first type of binary star to be observed was the \emph{spatially-resolved} class. These are also called visual binaries (because they can be identified by eye) and astrometric binaries (because it is possible to determine their orbits from measurements of the relative positions of the two stars).

Visual binaries were shown to be real, rather than chance alignments of stars at different distances, by the Revd.\ John Michell (\citeyear{Michell1767rspt}) on statistical grounds. William Herschel (\citeyear{Herschel1802rspt}) introduced the term \textit{binary star} and spent many years proving that some visual doubles show orbital motion. \citet{Herschel1803rspt} found that the double star Castor, with a separation of 3.9$^{\prime\prime}$, is a binary system with a period of 342\,yr, relatively close to the ``modern'' value of 420\,yr \citep{Rabe58an}. Visual binaries are usually close to Earth and have a large orbital separation (and thus a long orbital period) in order for the stars to be far enough apart on the sky to be resolved.

The equations of an astrometric orbit were established by F\'elix Savary in 1827. From observations of the motion of one of the stars relative to the other, it is possible to determine several of the properties of the orbit ($P$, $e$, $\omega$, $a$). The semimajor axis $a$ that can be found is the \emph{angular size}, not the true length (Table\,\ref{tab:par}). The properties $\omega$, $\Omega$ and $i$ can also be measured: these give the orientation of the orbit relative to the observer so are not intrinsic properties of the system.

If we know the distance to the system, usually via a parallax from the \textit{Hipparcos} or GAIA satellites, we can convert $a$ from angular units to length units. As this is the semimajor axis of the \emph{relative orbit}, it can be used to find the sum of the masses (via Kepler's Third Law) but not the individual masses.

If we obtain radial velocity (RV) observations of the two stars and fit a spectroscopic orbit (see below), we can convert $a$ from angular units to length units, measure the distance to the system (without needing its parallax) and also calculate the individual masses of the stars. We can therefore use these systems as distance indicators \citep[e.g.][]{Torres++97apj} and to constrain the predictions of theoretical models of stellar evolution \citep[e.g.][]{Torres++09apj}.

\begin{table} \centering
\setlength{\tabcolsep}{2pt}
\caption{\label{tab:par} Symbols and names of quantities measurable from various types of binary star.}
\begin{tabular}{|ll||c|c|c||c|c||c|c|c|}
\hline
                                  &                 &\multicolumn{3}{|c||}{Astrometric binary}&\multicolumn{2}{|c||}{Spectroscopic}&\multicolumn{3}{|c|}{Eclipsing binary}\\
\cline{3-5} \cline{8-10}
                                  &                 &             & with         & with       & \multicolumn{2}{|c||}{binary}      &            & with RVs    & with RVs  \\
\cline{6-7}
Name                              & Symbol          &   alone     & distance     & RVs        &     \,~SB1~\,     &        SB2     &   alone    & (SB1)       & (SB2)     \\
\hline \multicolumn{10}{l}{\textit{Orbital parameters}} \\ \hline
Orbital period                    & $P$             &      *      &       *      &      *     &         *         &         *      &     *      &      *      &      *    \\
Orbital eccentricity              & $e$             &      *      &       *      &      *     &         *         &         *      &     *      &      *      &      *    \\
Argument of periastron            & $\omega$        &      *      &       *      &      *     &         *         &         *      &     *      &      *      &      *    \\
Longitude of ascending node       & $\Omega$        &      *      &       *      &      *     &                   &                &            &             &           \\
Projected semimajor axis          & $a \sin i$      &             &       *      &      *     &                   &         *      &            &             &      *    \\
True semimajor axis               & $a$ (au)        &             &       *      &      *     &                   &                &            &             &      *    \\
Orbital inclination               & $i$             &      *      &       *      &      *     &                   &                &     *      &      *      &      *    \\
Distance                          & $d$             &             &              &      *     &                   &                &            &             &      *    \\
\hline \multicolumn{10}{l}{\textit{Spectroscopic parameters}} \\ \hline
Velocity amplitude of star 1      & $K_1$           &             &              &      *     &         *         &         *      &            &      *      &      *    \\
Velocity amplitude of star 2      & $K_2$           &             &              &      *     &                   &         *      &            &             &      *    \\
Systemic velocity                 & $V_\gamma$      &             &              &      *     &         *         &         *      &            &      *      &      *    \\
Mass function                     & $f(M)$          &             &              &            &         *         &         *      &            &      *      &      *    \\
Mass ratio                        & $q = M_2/M_1$   &             &              &      *     &                   &         *      &            &             &      *    \\
Mass sum                          & $M_1+M_2$       &             &       *      &      *     &                   &         *      &            &             &      *    \\
Minimum masses                    & $M_{1,2} \sin^3i$ &           &              &      *     &                   &         *      &            &             &      *    \\
Mass of primary star              & $M_1$           &             &              &      *     &                   &                &            &             &      *    \\
Mass of secondary star            & $M_2$           &             &              &      *     &                   &                &            &             &      *    \\
\hline \multicolumn{10}{l}{\textit{Size parameters}} \\ \hline
Fractional radii                  & $r_1$ and $r_2$ &             &              &            &                   &                &     *      &      *      &      *    \\
Radius of primary star            & $R_1$           &             &              &            &                   &                &            &             &      *    \\
Radius of secondary star          & $R_2$           &             &              &            &                   &                &            &             &      *    \\
Surface gravity of primary        & $\log g_1$      &             &              &            &                   &                &            &             &      *    \\
Surface gravity of secondary      & $\log g_2$      &             &              &            &                   &                &            &      *      &      *    \\
Density of primary star           & $\rho_1$        &             &              &            &                   &                &            &             &      *    \\
Density of secondary star         & $\rho_2$        &             &              &            &                   &                &            &             &      *    \\
\hline \multicolumn{10}{l}{\textit{Radiative parameters}} \\ \hline
Temperature of primary star       & $T_{\rm eff,1}$ &      *      &       *      &      *     &         *         &         *      &            &      *      &      *    \\
Temperature of secondary star     & $T_{\rm eff,2}$ &      *      &       *      &      *     &                   &         *      &            &             &      *    \\
Luminosity of primary star        & $L_1$           &             &              &      *     &                   &                &            &             &      *    \\
Luminosity of secondary star      & $L_2$           &             &              &      *     &                   &                &            &             &      *    \\
\hline
\end{tabular}
\end{table}

\subsection{Spectroscopic binaries}

Spectroscopic binaries are those found from changes in the RVs of the stars due to orbital motion. The first known spectroscopic binary was Algol ($\beta$\,Persei), which was already known to be an eclipsing binary (see below). Hermann \citet{Vogel1890pasp} observed the brighter of the two components to be moving away from Earth before primary eclipse and moving towards Earth after primary eclipse. Rudolf Lehmann-Filh\'es established the equations of a spectroscopic orbit in their current form in 1894.

For a ``single-lined'' spectroscopic binary system (SB1) -- one where we can measure RVs for only one component -- we can measure the quantities $P$, $e$, $\omega$, $K_1$ and $V_\gamma$ (Table\,\ref{tab:par}). From these we can in turn calculate the mass function $f(M)$ and a lower limit on the semimajor axis of the barycentric orbit of the star, $a_1\sin i$.

For a ``double-lined'' spectroscopic binary system (SB2) -- where RVs are measurable for both components -- we can obtain all the quantities for SB1 systems, plus the minimum masses ($M_1 \sin^3i$ and $M_2 \sin^3i$) and a lower limit on the semimajor axis ($a\sin i$). As the orbital inclination $i$ is not known, the true values of these three quantities are not accessible.

Spectroscopic binaries are easier to study if they have short orbital periods, large masses and a high $i$, in order to maximise the amplitude of the RV variation. On the other hand, it is more difficult to measure RVs for massive stars because they have few spectral lines and high rotational velocities \citep[see][]{MeClausen07aa}. Spectroscopic binaries are useful for measuring the multiplicity fraction of stars \citep[e.g.][]{DuquennoyMayor91aa}, which varies as a function of mass \citep{DucheneKraus13araa}, age \citep{Jaehnig+17apj} and chemical composition \citep{Badenes+18apj} and can be used to probe the star-formation process \citep[e.g.][]{Bate09mn}.

\subsection{Eclipsing binaries}

These are the most useful kind of binary star because of the huge number of physical properties that can be measured to high precision and accuracy (Table\,\ref{tab:par}). John \citet{Goodricke1783} is generally credited as the first to advance the hypothesis of eclipses, in order to explain the dimming of the ``demon star'' Algol every 2.87 days. These eclipses are even recorded in the Ancient Egyptian Calendar dating from around 1100\,B.C.\ \citep{JetsuPorceddu15plos}. The first eclipsing binary system to be properly characterised was $\beta$\,Aurigae by \citet{Stebbins11apj}, and the values found for its masses and radii agree reasonably well with modern values \citep{Me++07aa}. The mathematical framework for eclipse calculation was laid out by \citet{Russell12apj,Russell12apj2}; see also \citet{Kopal59book}.

The light curve of an eclipsing binary depends on the \emph{fractional radii} of the two stars ($r_1 = \frac{R_1}{a}$ and $\frac{R_2}{a}$), so cannot be used in isolation to obtain the full physical properties of the system (Table\,\ref{tab:par}). The orbital eccentricity $e$ and argument of periastron $\omega$ can also be found, because $e\cos\omega$ depends on the time of secondary eclipse relative to primary eclipse, and $e\sin\omega$ depends on the relative durations of the eclipses.

If we add RVs of one star to the light curve of an EB we can measure the mass function, $f(M)$, but not the actual masses of the stars. The surface gravity of the secondary star (the one for which we have no RVs) can also be calculated \citep{Me+04mn3}, something that can even be applied to transiting planets \citep{Me++07mn}.

If RVs are available for both stars then the full physical properties of the system can be obtained: the true masses and radii of both stars, which in turn give the surface gravities and densities. In the best cases these quantities can be measured to a precision of 0.5\% or better \citep[e.g.][]{Helminiak+19mn}. It is usually the case that the temperatures of the stars are known, from spectroscopy plus the surface brightness ratio measured from the light curve. In this case the luminosities of the stars can be calculated from $L = 4\pi R^2\sigma T_{\rm eff}^{~4}$, and the distance to the system can be obtained. EB-based distance estimates have been published for nearby galaxies: the LMC \citep{Pietrzynski+19nat}, SMC \citep{North++10aa} and M33 \citep{Bonanos+06apj}.

For the case of SB2 EBs we can measure the masses, radii and luminosities of two stars of the same age and chemical composition. Such information is valuable for calibrating theoretical stellar models \citep[e.g.][]{Andersen++90apj,Pols+97mn,Claret07aa}, if the two stars have not undergone mass transfer and thus have evolved as single stars. \citet{ClaretTorres16aa,ClaretTorres18apj} have used a large sample of SB2 EBs to calibrate the strength of convective core overshooting as a function of mass, finding a ramp-up from 1.2\,M$_\odot$ to 2.0\,M$_\odot$. However, this approach has been challenged by \citet{Valle+16aa} due to possible biases in the method, and by \citet{ConstantinoBaraffe18aa} about the reproducibility of the results.

It is possible to add interferometry of SB2 EBs and obtain highly precise distances and physical properties of individual systems \citep{Gallenne+16aa,Gallenne+19xxx}. The study of EBs with pulsating components is a very promising possibility for constraining the interior and atmospheric structure of stars \citep{Debosscher+13aa,Aerts13conf,Tkachenko+14mn,Themessl+18mn}. Catalogues of EBs from photometric surveys are useful for determining the multiplicity fraction of stars, as have been achieved for M-dwarfs by \citet{Shan++15apj} and for solar-type stars by \citet{Moe++19apj}.

\subsubsection{DEBCat: the Detached Eclipsing Binary Catalogue}

DEBCat\footnote{\texttt{https://www.astro.keele.ac.uk/jkt/debcat/}} \citep{Me15debcat} is a catalogue of EBs suitable for comparison with predictions from theoretical stellar models. It includes all EBs for which there is no evidence for current or past mass transfer, and with mass and radius measurements to accuracies of 2\% or better. The full physical properties of the systems are collected (mass, radius, surface gravity, temperature, luminosity, orbital period, metallicity) and can be downloaded in ascii format. At the time of writing (2019/12/31) DEBCat contains 239 systems.

\section{1SWASP J034114.25+201253.5}

\begin{figure}[t] \centering
\includegraphics[width=\textwidth,clip]{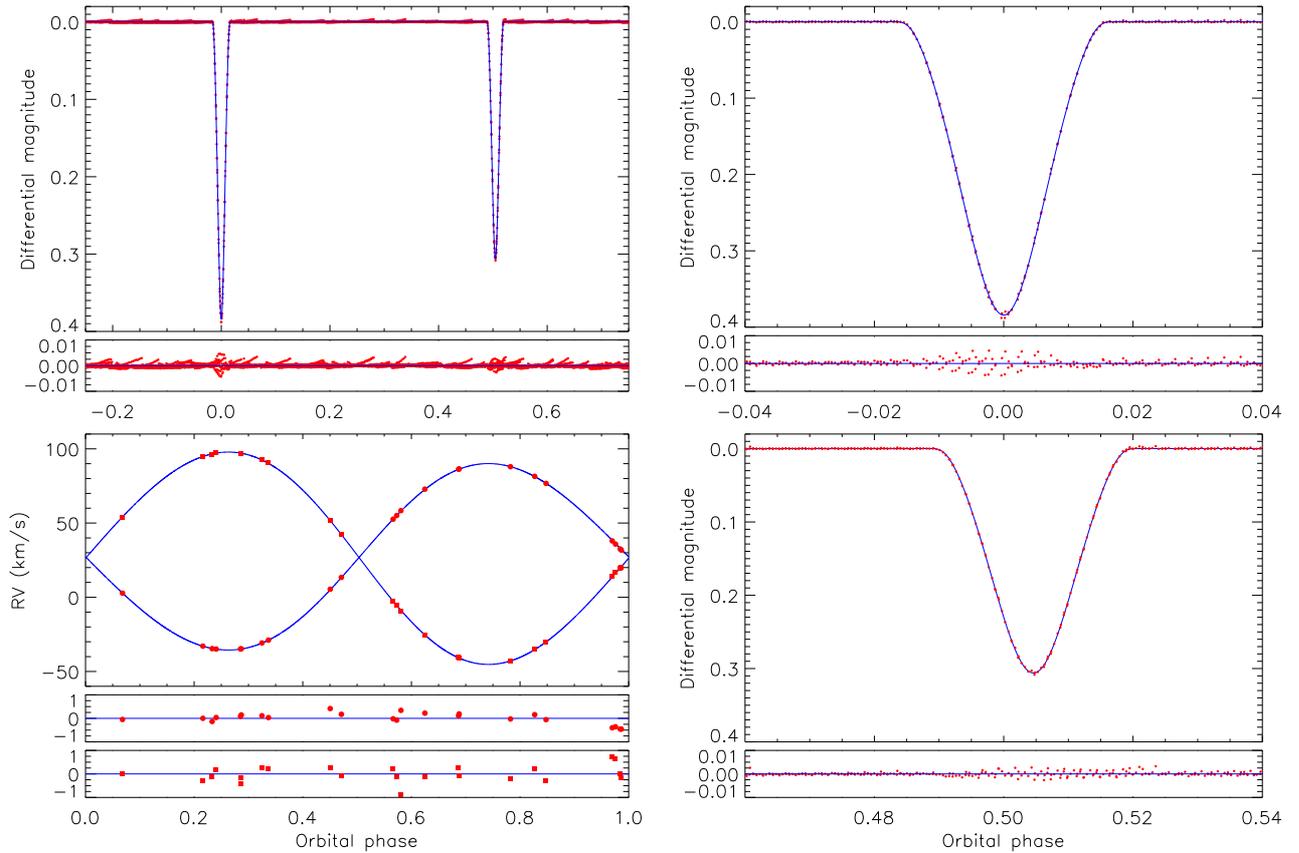}
\caption{\label{fig:0341} Observational data for WASP\,J0341.
Top left: K2 light curve versus orbital phase.
Top right: primary eclipse.
Bottom right: secondary eclipse.
Bottom left: RV curves.}
\end{figure}

I now present some preliminary work on the solar-type eclipsing binary system 1SWASP J034114.25+201253.5 (hereafter WASP\,J0341), performed in collaboration with P.\ Maxted (Keele), G.\ Torres (CfA) and K.\ Pavlovski (Zagreb). WASP\,J0341 was identified as an eclipsing binary in the vicinity of the Pleiades open cluster using photometric data from the SuperWASP database \citep{Pollacco+06pasp}. Because of the scientific importance of eclipsing binaries that are cluster members \citep[e.g.][]{Me++04mn,Brogaard+11aa} we proceeded to obtain several sets of additional observations. We applied for and were awarded observations of WASP\,J0341 from the NASA K2 mission, which observed it in long cadence for 71 days during Campaign 4. The PDC light curve from the \textit{Kepler} data reduction pipeline is shown in Fig.\,\ref{fig:0341}; more sophisticated reductions of these data exist and will be used for future analyses.

We also obtained a set of 30 high-resolution spectra of this system from the 1.5\,m TRES spectrograph on the 1.5\,m Tillinghast telescope at FLWO. RVs from these data are shown in Fig.\,\ref{fig:0341}. A joint fit of these RVs and the K2 light curve using the {\sc jktebop} code \citep{Me13aa} yields masses of 1.08 and 0.95 M$_\odot$ and radii of 1.21 and 0.93 R$_\odot$, all measured to a precision of 0.5\% or better. We have also obtained a set of six high-S/N spectra from VLT/UVES that will be used to measure the chemical composition of the two stars, yielding an even more precise test of theoretical models.

WASP\,J0341 has an 8-day orbital period and a small eccentricity, so tidal effects in this system are small. Its properties will therefore serve as an excellent test of theoretical models of solar-type stars. A careful inspection of Fig.\,\ref{fig:0341} shows that the scatter of the data increases during both eclipses. This is an indication of temporal and spatial changes in the flux from the stellar surfaces, which is likely due to starspots. Our UVES data do show weak emission in the centres of the Ca H \& K lines, a good indicator of magnetic activity in the stars \citep{Baliunas+95apj}.

\section{Conclusions}

Binary system are a common result of the star formation process and offer the only known way of determining the masses of stars directly. Their frequency of occurrence offers a way to probe the star formation process. Eclipsing binary systems are those which can be characterised in detail: precise measurements of their properties allows an exacting assessment of the predictive power of theoretical stellar models and provides one of the lower rungs of the cosmological distance ladder.

The advent of space satellites has revolutionised the study of eclipsing binaries, with the discovery of several thousand examples using CoRoT \citep{Deleuil+18aa} and \textit{Kepler} \citep{Kirk+16aj}. The TESS satellite is currently providing new data on the great majority of known eclipsing binaries, and in future the PLATO mission will provide unparalleled photometric observations of many known and new examples.

% Optional acknowledgements
% -------------------------
\begin{acknowledgements}
I thank the organisers of the conference for their invitation to give this review talk, my many collaborators on eclipsing binaries, and the students of the PHY-30024 module who are gradually and uncomplainingly weeding out all the typographical and mathematical errors in my lecture notes on binary stars and extrasolar planets.
\end{acknowledgements}

%%-----------------------------
%%   Bibliography
%%-----------------------------
%%
%% The format for references is the one adopted by A&A (see the example below).
%% To set the reference list in the proper format, we ask you to use BibTEX and %% the natbib package instead of the standard 'thebibliography' environment.
%%
%% The following lines are then required:
\bibliographystyle{aa}  % A&A bibliography style file (aa.bst)
% \bibliography{southworth_6k01} % your references in file: surname_nxnn.bib
% \bibliography{aamnem99,jkt}

%
\end{document}